\documentstyle[mncite]{mn} 

\def\nice#1::::{#1}    \def\subm#1::::{}   

\def\dosingle#1::::{#1}  \def\dodouble#1::::{ } 
\newcommand\zzz[2]{#2}  

\def\mod{\mbox{\rm mod}}

\def\k{\mbox{\rm\,km\,s$^{-1}$}}

\def\centreline{\centerline}
\def\.{{\cdot}}
\def\gtapprox{\,\lower.6ex\hbox{$\buildrel >\over \sim$} \, }
\def\ltapprox{\,\lower.6ex\hbox{$\buildrel <\over \sim$} \, }
\def\sun{\odot}

\def\e{ {\scriptstyle \times} 10^}

\def\arcs{\ifmmode {'' }\else $'' $\fi}     
\def\arcm{\ifmmode {' }\else $' $\fi}     
\def\deg{\ifmmode^\circ\else$^\circ$\fi}    

\def\rComa{{\bf r}_{\mbox{\rm \small Coma}}}  

\def\apj{Ap.J.}                 
\def\apjs{Ap.J.Supp.}                 
\def\aj{A.J.}                       
\def\aanda{A.\&A.}            
\def\frtoday{le\space\number\day\space\ifcase\month\or
  janvier\or f\'evrier\or mars\or avril\or mai\or juin\or
  juillet\or ao\^ut\or septembre\or octobre\or novembre\or d\'ecembre\fi\space \number\year}

\nice \input epsf ::::

\dodouble \documentstyle[doublespacing,mncite]{mn} ::::

\begin{document}

\newcommand\joref[5]{#1, #5, {#2, }{#3, } #4}
\newcommand\inpress[5]{#1, #5, #4}
\newcommand\epref[3]{#1, #3, #2}

\def\jref#1;;#2;;#3;;#4 {#1, {#2, }{#3, }#4}

\def\bref#1;;#2;;#3;;#4;;#5 {#1, {in #2 }(#3: #4), #5}
\def\apjpre#1;; {#1, preprint}

\def\apjpriv#1;; {#1, private communication}

\def\apjprpn#1;; {#1, in preparation}

\def\apjsub#1;;#2;; {#1, #2, submitted}

\def\apjprss#1;;#2;; {#1, #2, in press}

\def\naoj{National Astronomical Observatory, Mitaka, Tokyo 181, Japan}
\def\IAP{Institut d'Astrophysique de Paris, 98bis Bd Arago, F-75.014 Paris,
France}
\def\sussex{Astronomy Centre, 
MAPS, University of Sussex, Falmer, Brighton,
BN1~9QH, United Kingdom}
\def\ioa{Institute of Astronomy, University of Cambridge, Madingley Road, CB3 OHA, United Kingdom}
\def\junk{lots and lots of ze words;  }

\title[X-ray Cluster Topology Constraint]{Constraining Cosmological Topology via Highly Luminous X-ray Clusters}

\author[B.F.~Roukema \& A.C.~Edge]{Boudewijn F.~Roukema$^1$ and Alastair C.~Edge$^2$\\ {$^1$\naoj}\\ {$^2$\ioa} }

\def\today{\frtoday}

\def\oo{$\ddot{\mbox{\rm o}}$}
\def\LaLu{Lachi\`eze-Rey \& Luminet (1995)}

\maketitle

\newfont{\sans}{cmss10}


\begin{abstract}
The topology of the observable Universe is not yet known. The most 
significant observational sign of a non-trivial 
topology would be multiple images (``ghosts'') 
of a single object 
at (in general) different sky positions and redshifts. 

It is pointed out that the previous search by \cite{Gott80} (1980) for 
ghost images of the Coma cluster can be extended by using 
highly X-ray luminous clusters of galaxies. 
This is likely to be more efficient than with other astrophysical objects 
viewable on these scales 
since (1) X-ray clusters would be at least as 
easy to identify if viewed from other angles as any other objects 
and (2) the X-ray emitting thermally 
heated gas is likely to be simpler than for other objects.

Possibilities that the highly luminous cluster 
RXJ~1347.5-1145 ($z=0\.45$) has a ``ghost image'' at lower redshift 
are analysed. 
It is noted that 
RXJ~1347.5-1145, the Coma cluster and the cluster CL~09104+4109 
form nearly a right angle ($\approx 88\deg$) with arms of nearly identical 
length ($970h^{-1}$ and $960h^{-1}$~Mpc respectively) for 
$\Omega_0=1, \lambda_0=0$ curvature 
($h\equiv H_0/100$km~s$^{-1}$~Mpc$^{-1}$). This is a clue 
that the three clusters could be ghost images of 
one and the same cluster, 
for a hypertoroidal topology. However, several arguments are 
presented that this relation is not physical. 
\end{abstract}

\begin{keywords}
methods: observational --- 
cosmology: observations --- galaxies: clusters: individual (RXJ~1347.5-1145) --- galaxies: clusters: individual (Coma) --- X-rays: galaxies 
\end{keywords}

\def\tabone{
\begin{table*} 
\caption{\label{t-clusters} Basic data on highly luminous X-ray clusters, 
assuming $\Omega_0=1, \lambda_0=0, H_0=50$~kms$^{-1}$~Mpc$^{-1}$. 
$M_r$ indicates total mass to $r$~Mpc in units of
$10^{14}M_{\sun}$; $L_X($W$)$ 
is X-ray luminosity in units of $10^{44}$~erg/s in band W=1 
($0\.1-2\.4$~keV), W=2 (2-10~keV) or W=b ({\rm ``bolometric''}),
$T, r$ is temperature $T$ in keV within a radius $r$. 
Luminosities for A2163, Coma, A1835 and Perseus are from
\protect\cite{Ebel96a}~(1996a); other 
references are \protect\cite{Elbaz95}~(1995), 
\protect\cite{Sch96}~(1996),
\protect\cite{Briel92}~(1992),
\protect\cite{Hall97}~(1997), \protect\cite{FabCr95}~(1995) and
\protect\cite{Allen96}~(1996)
as indicated by abbreviations. }
$$\begin{array}{l l r c@{\ \ \ }c c c c ccc c c}
\hline 
\multicolumn{1}{c}{{\rm Object}} & \alpha & \delta & z 
& M_{1\.5} & M_{3}& M_{4\.6} & M_5 
& L_X(1)  & L_X(2)  &  
L_X({\rm b}) & T, r & {\rm ref.}\\             
\hline 
\mbox{\rule[-0ex]{0cm}{2ex}}  
{\rm A2163} & 16{}^h{}15{}^m{}\.75{} & -6{}\deg{}09{}\arcm{} & 0\.201 
& & &46^{+4}_{-15} & 
& 38 & 41 &  & 14\.6\pm1, 4\.2\arcm{}
& {\mbox{\rm Elb95}} \\
{\rm RXJ~1347.5-1145}& 13{}^h{}47{}^m{}\.5{}& -11{}\deg{}45{}\arcm{}& 0\.451 
& &17 & &   
& 73 &76 &200 & 9\.3\pm1, 7\arcm{} 
& {\mbox{\rm Sch96}} \\
{\rm Coma \ cluster} & 12{}^h{}57{}^m{}{} & 28{}\deg{}15{}\arcm{} & 0\.023
& &15\pm5 & &18\pm6  
& 7\.2 &8\.8 &  & 8\pm3, 16\arcm{}
& {\mbox{\rm Br92}} \\
{\rm CL~09104+4109} & 09{}^h{}10{}^m{}\.5{} & 41{}\deg{}09{}\arcm{} & 0\.442
& & & & & 30 
& & & 11\.4^{+\infty}_{-3.2}
& {\mbox{\rm H97, FC95}} \\
{\rm CL~1821+643} & 18{}^h{}21{}^m{}{} & 64{}\deg{}03{}\arcm{} & 0\.297 
& & & & & 37
& & &  
& {\mbox{\rm H97}} \\
{\rm A1835} & 14{}^h{}01{}^m{}{} & 02{}\deg{}53{}\arcm{} & 0\.252 
& 9 & & & 
& 38 & 43 & & 9\.5\pm2, 6\arcm{} 
& {\mbox{\rm A97}} \\
{\rm Perseus \ cluster} & 03{}^h{}15{}^m{}\.3{} & 41{}\deg{}20{}\arcm{} 
& 0\.018
 &&&& &
12\.7 & 11\.9 & & 6\.7\pm2 \\
\hline
\end{array} $$
\end{table*}
} 

\def\tabbas{
\begin{table} 
\caption{\label{t-basis} Basis vectors  $L{\bf e}_i$ 
of candidate manifold, in cartesian 
equatorial coordinates ($x= r \cos \delta\,\cos\alpha,$ 
$y= r \cos \delta \,\sin \alpha,$
$z= r \sin \delta,$ for radial comoving distance $r$) in units of 
$h^{-1}$~Mpc. $\Omega_0=1, \lambda_0=0$ is assumed. The objects defining
these basis vectors are listed (but note that the vectors are modified in
order to give an exact cube as the fundamental polyhedron, of 
side length $L=962h^{-1}$~Mpc).
}
$$\begin{array}{l c c c c}
\mbox{\rm Motivation for } L{\bf e}_i  & 
L{\bf e}_i &x_i & y_i & z_i \\
\hline 
\mbox{\rule[-0ex]{0cm}{2ex}}  
\mbox{\rm Coma to RXJ~1347.5-1145} & L{\bf e}_1 & -813  &  -446  &  -254 \\
\mbox{\rm Coma to CL~09104+4109} &  L{\bf e}_2 & -490  &   533   &  633\\
\mbox{\rm orthonorm. to vectors 1, 2} & L{\bf e}_3 &-153  &   665  &  -678\\
\hline
\end{array} $$
\end{table}
} 

\def\FigLxplot{ 
\begin{figure} 
\nice \centreline{\epsfxsize=6.5cm
\zzz{\epsfbox[82 45 480 600]{"`gunzip -c Lxplot.ps.gz"} }
{\epsfbox[82 45 480 600]{"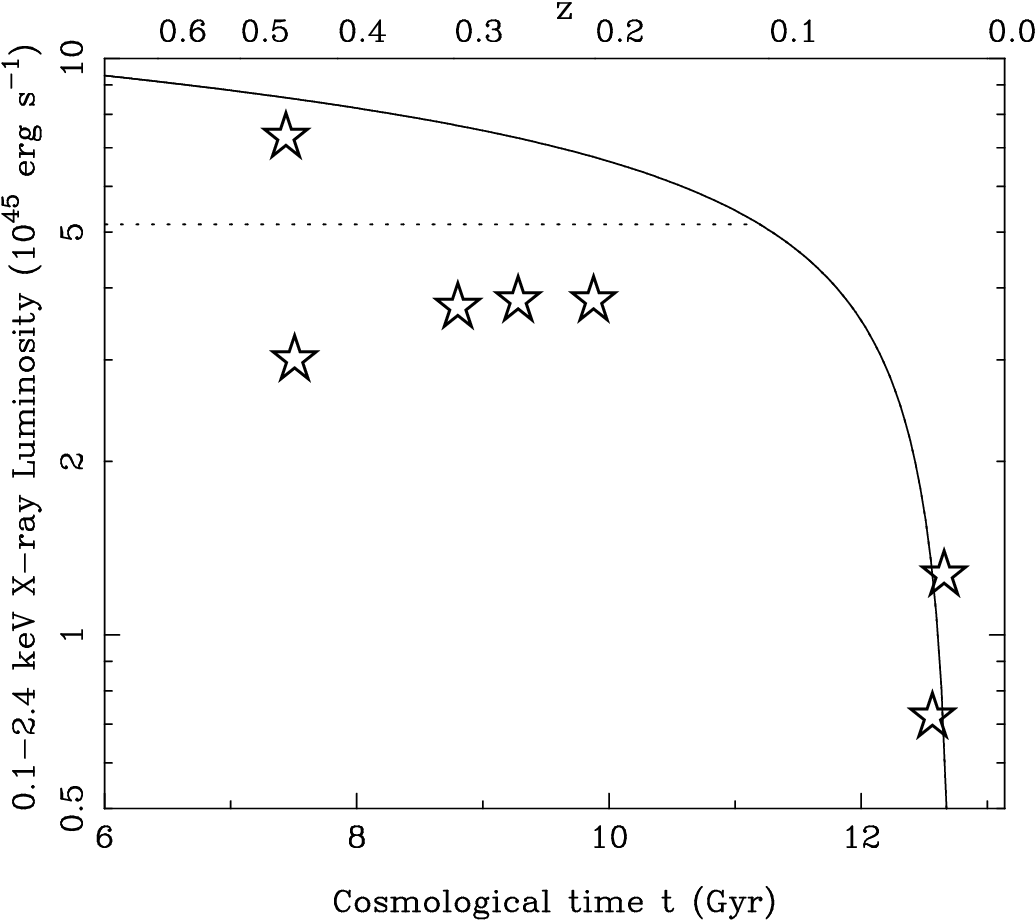"} }
}
 ::::
\subm \vspace{5cm} ::::
\caption{\label{f-Lxplot} \protect\footnotesize
X-ray luminosities of several of the brightest
clusters shown against cosmological time (values in 
Table~\protect\ref{t-clusters}). A solid curve shows the brightest
luminosity at redshift $z$ to which the observed $0\.1-2\.4$~keV X-ray 
luminosity function (parametrised as a Schechter function; 
values and units of \protect\cite{Ebel97}~1997) 
predicts a number density of one object 
per the total observable volume to $z$, assuming a trivial topology. 
(One-third of the sky is assumed 
unobservable due to the galactic plane.) The dotted curve shows the
same brightest luminosity statistic for a universe of non-trivial
topology, with a fundamental polyhedron diameter of 600$h^{-1}$~Mpc.
($\Omega_0=1, \lambda_0=0, h=0\.5$ are used.)
}
\end{figure} 
}

\def\FigQobs{ 
\begin{figure}  
\nice \centreline{\epsfxsize=8cm
\zzz{ \epsfbox[32 25 430 725]{"`gunzip -c hypobs.ps.gz"} }
{ \epsfbox[32 25 430 725]{"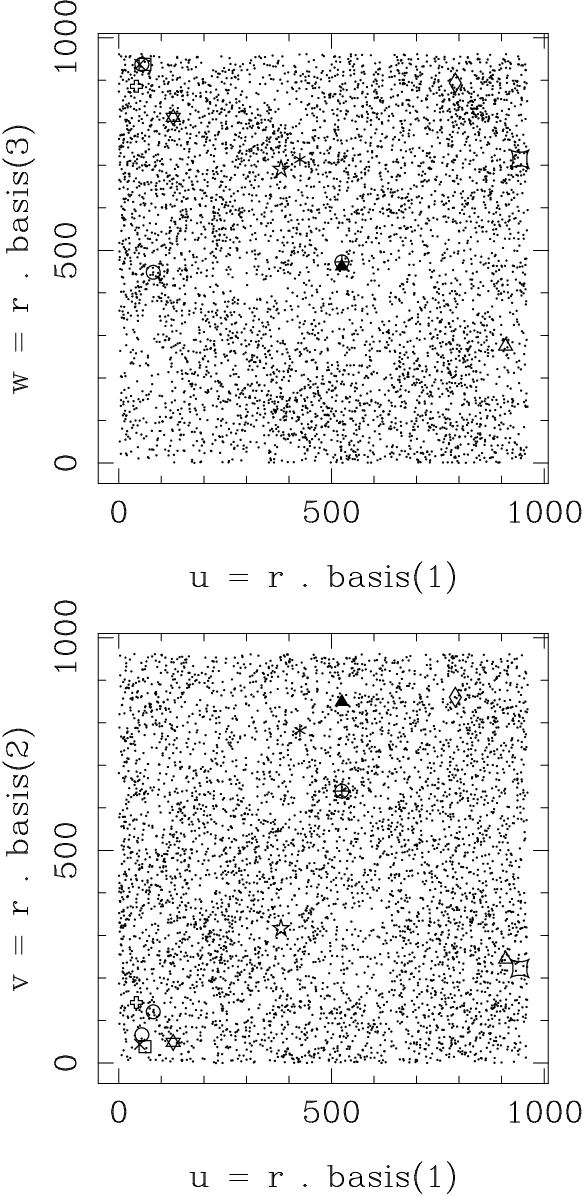"} }
}
 ::::
\subm \vspace{9cm} ::::
\caption{\label{f-qobs} \protect\footnotesize
Observed quasars (N=5007) shifted to fundamental cube according to the
linear transformation defined by 
identifying RXJ~1347.5-1145, the Coma cluster and the cluster CL~09104+4109 
and supposing the third axis to be perpendicular and of the same
length as these identities (Table~\protect\ref{t-basis}; 
Eqn~\protect\ref{e-hyper}).
The points represent quasars; higher quality 
topological standard candles plotted are 
clusters 
A2163 (asterisk), RXJ~1347.5 (circle), 
Coma (``x''), CL~09104+4109 (square),
     CL~1821+643 (triangle); superclusters Ursa Major (Swiss cross) and 
CrB (Star of David); and six 
CMB cold spots (other large symbols). The three clusters defining the
transformation can be seen lying nearly on top of one another in the 
bottom-left of the $u-v$ panel and the top-left of the $u-w$ panel. 
}
\end{figure} 
}

\def\FigQsim{ 
\begin{figure}  
\nice \centreline{\epsfxsize=8cm
\zzz{\epsfbox[32 25 430 725]{"`gunzip -c hyprot.ps.gz"} } 
{\epsfbox[32 25 430 725]{"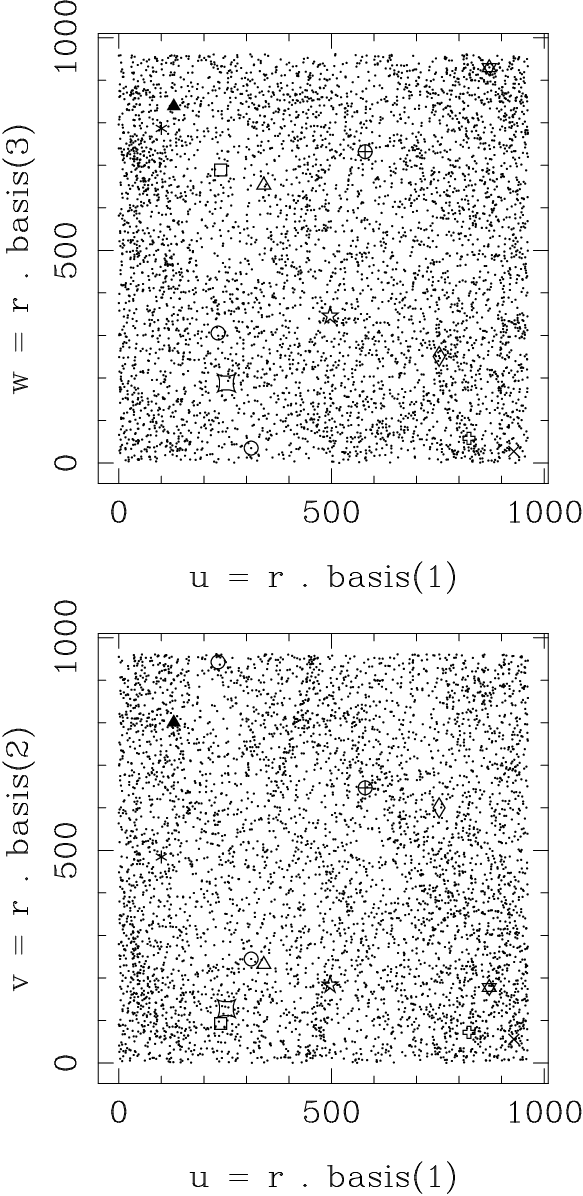"} } 
}
::::
\subm \vspace{9cm} ::::
\caption{\label{f-qsim} \protect\footnotesize
Same as for Fig.~\protect\ref{f-qobs}, except that the transformation
has been rotated by an arbitrary angle, so that coincidences between
objects should not be due to non-trivial topology.
}
\end{figure} 
}

\dodouble \clearpage ::::

\section{Introduction}
	From the beginning of modern cosmology, it has been realised that
both the curvature and topology of the observable Universe may be
nontrivial (e.g., \cite{deSitt17}~1917; \cite{Lemait58}~1958),
but the fact that objects at cosmological distances
are seen in our past time cone has meant that constraints on topology 
suffer astrophysical complications just as do the constraints on curvature
and on the Hubble constant. 

	The empirical tradition in cosmology requires that observations are
used to constrain both curvature and topology. Theoretical
indications that the topology should be nontrivial are 
not lacking, however. Indeed, observational 
indications are that the curvature of the Universe
is either negative or zero, implying that for a trivial topology the 
volume would be infinite. This implies that the requirement
of quantum gravity with a ``no-boundary'' boundary condition 
(e.g., Hawking 1984; Zel'dovich \& Grishchuk 1984) 
that the Universe be compact 
is not satisfied by a trivial
(simply connected) topology. A multi-connected negatively curved (or flat) 
universe of finite volume provides such compactness.

A completely independent argument for compactness is
from unified particle physics theories which imply
that most of 10 or 11 dimensions of the fundamental ``strings'' are 
``compactified''. If the volume of the Universe were infinite, it would
seem somewhat arbitrary that the three space dimensions are an exception
to the compactification of the other dimensions.
(This argument is known
as ``Dimensional Democracy, but where some dimensions are more equal
than others'', \cite{Stark96}~1996; \cite{Orwell}~1945.) Multi-connectedness
can save a finite, non-positively curved universe. 

A third argument comes from inflationary theory for negatively curved
universes. Observational difficulties with an $\Omega_0=1$ universe
have stimulated inflation models which predict that 
$\Omega_0+\lambda_0<1$ (\cite{Gott80}~1982, 1986; 
\cite{Sasaki93}~1993; \cite{Linde95}~1995; for consequences on the 
cosmic microwave background, 
see \cite{RaPee94}~1994; \cite{BuGo95}~1995).
However, inflation resulting in negative curvature at the present epoch
requires fine-tuning the initial conditions, which is provided 
in these models by an earlier period of inflation. In that case the
inflationary potential is designed in order to yield such 
``double inflation''. On the other hand, 
multi-connectedness in a negatively curved universe implies chaoticity
of geodesics, and so naturally provides the smooth initial conditions
for a single period of inflation which
results in observable curvature at the present (\cite{Corn96}~1996).

Lachi\`eze-Rey \& Luminet (1995) present an extensive introductory review of 
how one can understand non-trivial topology in a standard hot big, bang
(Friedmann-Lema\^{\i}tre-Robertson-Walker) Universe, examples of 
many of the known (orientable) topologies possible, and of observational
efforts to measure or constrain the topology. 

However, the reader is reminded here of the simplest example of a 
three-dimensional topology, for trivial curvature---the 
hypertorus. One can think of the entire
Universe as a cube, of side length at least several hundred or thousand Mpc, 
of which opposite faces are physically identified. 
A particle such as a photon may travel between two points in space, 
e.g., from object to observer, by
a direct route ``within'' the cube, or may ``pass through'' the faces of
the cube several times in travelling from one point to the other. 
Hence, many ``ghost'' images at (in general) differing celestial positions and
(in general) differing redshifts (due to differing light travel time) 
would be seen. Of course, in proper coordinates, this cube expands in
size according to the scale factor, so that detection of non-trivial
topology has to be done in comoving coordinates.

This embedding within infinite, flat, three-dimensional space is only an
aid to our intuition---the space outside the cube 
does not have physical significance, although it
is useful mathematically. The infinite 
space (or a hyperbolic or elliptic infinite
space for other curvatures) is termed the ``universal covering space''; 
the cube (or in other cases other polyhedra of which faces are identified
in some way) is termed the ``fundamental polyhedron''.
Following \cite{LaLu95}~1995, 
the shortest distance from an object to any of its ``ghost'' 
counterparts is $\alpha$, and the largest distance from an object to an 
adjacent ghost (where there is an adjacent ghost for each 
face of the fundamental polyhedron) is $\beta$. 

The most direct way of attempting to detect the topology of the Universe
is by searching for individual 
ghost images of known objects in the universal covering space. 
That is, objects observed in the past time cone are assumed to 
exist in a simply-connected universe, at distances they would have
(assuming zero peculiar velocity) within the simply-connected spatial 
hypersurface at $t=t_0$. This covering space would contain many copies of the
fundamental polyhedron. Constraints obtained by this technique 
(e.g., \cite{SokShv74}~1974; \cite{Gott80}~1980; 
\cite{Fag85}~1985, 1989; \cite{Fag87}~1987) give constraints on $\alpha, 
\beta$
to a few tens or hundreds of Mpc or else find ghost image candidates 
statistically consistent with chance coincidences.

Several recent attempts have been made 
to use the essentially two-dimensional slice of
the covering space which temporally corresponds to 
the period of recombination, i.e., which spatially is the surface of
last scattering observed as the cosmic
microwave background (CMB) by the COBE satellite 
(\cite{Stev93}~1993; \cite{Star93}~1993; 
\cite{JFang94}~1994; \cite{deOliv95}~1995). 
These methods rely upon our present understanding of
the local physics at the redshift of last scattering and require 
various assumptions, e.g., assume a toroidal topology, and claim limits
to a few thousand Mpc. An interesting advance in techniques of 
finding candidate manifolds, of hyperbolic (negative) curvature,
is that of \cite{Fag96}~(1996), in which possible identifications between
CMB cold spots and galaxy superclusters are used to predict expected 
positions ($\alpha, \delta, z$) of further ghosts of superclusters. 
Observational confirmation or refutation of such candidate manifolds
is obviously straightforward.

Three topology-independent methods either for three-dimensional 
catalogues of objects spread
over large scales (\cite{LLL96}~1996; \cite{Rouk96}~1996) or for the
CMB (\cite{Corn97}~1997) now provide alternatives to the search for 
individual ghosts and the CMB methods cited above.

However, the subject of this paper is an extension of the traditional
and simplest method, the search for ghost images of individual objects.
The explanation of why X-ray clusters are 
probably the best ``topology standard
candle'' available on scales of about $1000h^{-1}$~Mpc is presented
in \S\ref{s-Xray}, cases of specific clusters are discussed in 
\S\ref{s-individ} and a summary is given in \S\ref{s-concl}.

For brevity of discussion, and because of the theoretical motivations
for a finite volume universe,
the use below of the 
terms ``multi-connected'' and ``non-trivial topology'' should be
taken to imply the finite volume cases only unless otherwise specified. 

For reference, the reader should be reminded that 
the horizon is at $6000 h^{-1}$~Mpc from the observer, and
the horizon diameter is $12000 h^{-1}$~Mpc. 
(Except where otherwise stated, distances are quoted 
in comoving units in an $\Omega_0=1, \lambda_0=0$ universe 
and $h\equiv H_0/100$km~s$^{-1}$~Mpc$^{-1}$ is explicitly indicated;
cluster luminosities and masses implicitly include the assumptions
that $\Omega_0=1, \lambda_0=0, h=0\.5.$)

\section{X-ray Clusters as Topology Standard Candles} 
\label{s-Xray}
The Einstein, ASCA and ROSAT satellites have shown that many galaxy
clusters contain hot, bremsstrahlung, X-ray emitting gas. 
This is in fact the case for nearly all of 
the richest (Abell richness class $R >2$) clusters, and those not detected
may be undetected only due to the flux limit of the surveys 
(\cite{Ebel93}~1993). 

The richest clusters therefore provide what is 
probably the best ``topology standard
candle'' available on scales of about $1000h^{-1}$~Mpc:
(1) clusters should look 
fairly similar---in X-rays or optically in the statistical 
distribution of galaxies---from any direction;
(2) once the (rich) cluster exists, 
evolution of the X-ray emitting gas is likely to be only possible in 
one direction---to 
greater mass and/or greater mass concentration---the 
X-ray luminosity could probably not decrease by more than about a factor
of two.

These two properties are ideal for searching for ghosts. Property (1), 
isotropy, is useful since except in special
cases, ghosts are likely to be viewed from different angles, so the main
objects visible at large redshifts, quasars, are going to be much fainter
when seen from many angles (according to unified models
of active galactic nuclei they would be seen as, e.g., Seyfert galaxies). 
This is why \cite{LLL96}'s~(1996) statistical 
method 
is unlikely to be useful for quasars, so \cite{Rouk96}'s~(1996) method 
of searching for individual configurations was developed.

Property (2), the evolutionary constraint, is equally useful. Up to the
redshift of rich cluster formation, probably $z\gtapprox 1,$ the rarity
of the largest clusters in a simply-connected universe should imply that
surveys to larger and larger volumes should yield successively
rarer and higher mass clusters, simply since the higher the mass of a 
cluster the rarer it should be. 

\tabone

On the other hand, a multi-connected
universe of finite size smaller than the observed volume must contain a
cluster of maximum mass (for precise enough bins in mass determination), 
which forcibly has at least one ghost image in the copy of the fundamental
polyhedron which includes the observer. 
Ghosts of the maximum mass cluster in copies of the polyhedron
at higher redshifts are likely to 
be of identical or lower mass, if it is
accepted that the cluster can only gain in mass once formed. 

Hence, in a simply-connected universe, more and more massive (and rare) 
clusters should be seen to 
successively larger redshifts (up to the formation epoch of such clusters),
while in a small enough (finite volume) multi-connected universe, 
once {\em the} most massive cluster has
been seen, no clusters of greater mass should be seen at higher redshifts.

In other words, 
property (2) not only implies that X-ray emitting (rich) clusters
would in principle be possible to see as ghost images, but the existence 
at a ``high'' redshift 
of a single such cluster 
which is much more massive than all the others at lower redshifts
provides a strong constraint {\em against} multi-connectedness up to 
the redshift of the cluster. The key elements of this 
constraint are merely gravity and conservation of mass.

\subsection{Evolution} \label{s-evolution}
What are the possibilities for evolution of the gas component, which 
generates the X-ray luminosity?

This is likely to depend mostly on the mass of the cluster.
Observational links between the X-ray luminosity
of a cluster, $L_X,$ and the gas temperature, $T,$ 
(e.g., \cite{EdgeSt91}~1991; \cite{HenAr91}~1991), combined with
a theoretical isothermal potential $M-T$ relation (where $M$ 
is the virial mass; e.g., \cite{Evr96}~1996), 
indicate that $L_X \sim M^2.$
While this relationship is not evolutionary, it at least suggests that
if the mass of a cluster changes, its 
luminosity is likely to increase strongly.

Direct attempts to measure $L_X$ evolution with redshift $z$ have had diverse 
results (\cite{Gioia90}~1990; \cite{Edge90}~1990; \cite{Henry92}~1992;
\cite{LuppG95}~1995; \cite{OuBB97}~1997), but
\cite{Ebel96b}~(1996b; 1997) point out that several
of the analyses may have suffered selection effects, and find 
no significant evolution for 
X-ray selected ROSAT clusters to $z\ltapprox 0\.3.$
Even if the suggestions of ``negative'' evolution were correct,
this would mean that $L_X$ increases with time, strengthening
the usefulness of clusters as topological standard candles.

These results seem reasonable theoretically.
Since the gas is a much larger component (by mass) than the galaxies,
and given the time, length and mass scales involved\footnote{Remember 
that 1000\k $\approx 1$Mpc\,Gyr$^{-1}.$}, there is little possibility that
the hot gas can collapse mostly into galaxies (or $H_2$ clouds)
or be blown out of the cluster over several Gigayears. So the gas must
remain in the cluster, and since, at least to a good first approximation
(apart from cooling flows),
is in isothermal equilibrium in the potential well of the cluster, the
density distribution is unlikely to change by much. Since 
the X-ray emission
is believed to be bremsstrahlung emission due to particles accelerating in 
a potential well of $\sim10^{14-15}M_{\sun},$ it seems physically unlikely
that this emission could be suppressed by large factors.

The most physically reasonable
scenarios for evolution of the gas distribution (and the overall mass 
distribution) would be for the density distribution 
to become more centrally concentrated or for the mass to increase somewhat
due to gas (and galaxies) still infalling into the cluster. These cases
would agree with the EMSS data analysis
(\cite{Gioia94}~1994; \cite{OukB97}~1997) suggesting increase of $L_X$ with
time.
Interaction
with another large cluster might possibly loosen up the central concentration
(the core), but such a case is likely to be both rare and easy to spot.

Of course, a significant fraction of cluster X-ray luminosity often
comes from what are (usually believed to be) cooling flows
(e.g., \cite{FabCr95}~1995). Over a Hubble
time a cooling flow 
could feed a few percent of the cluster mass into the centre
of the cluster, not significantly changing the (approximately) 
isothermal gas distribution. If a cooling flow ceased between the 
epochs of two ghost images, the X-ray luminosity could decrease by
as much as a factor of two, lowering the chances of detection above
a given flux limit. However, this would again (by energetic 
considerations) require collision with another large cluster.

So, it is difficult to see how ghosts of ``high redshift'' 
X-ray clusters at lower redshifts 
could be too faint to have been detected. 

\FigLxplot

\section{Does RXJ~1347.5-1145 Have Any Ghost Images?} \label{s-individ}
Observational (or observationally deduced) parameters of RXJ~1347.5-1145 
 \cite{Sch96}~(1996) 
and of some of the most highly luminous X-ray clusters which could be
candidates for a ghost image of RXJ~1347.5-1145 are listed in 
Table~\ref{t-clusters}.

These luminosities (in the $0\.1-2\.4$~keV band, as illustration) are
shown graphically in in Fig.~\ref{f-Lxplot}. RXJ~1347.5-1145 is 
clearly exceptional relative to the other massive clusters.

The argument presented
in \S\ref{s-Xray} that brighter (or more massive), hence, rarer clusters 
should be seen to higher and higher redshifts (up to the cluster formation
epoch, and assuming little evolution) is also illustrated in this Figure.
From the \cite{Sch76}~(1976) function
fit to the ROSAT Brightest Cluster Sample of \cite{Ebel97}~(1997) it is 
straightforward for a given redshift to predict the luminosity such that
clusters of that luminosity become ``common'' enough (one occurrence)
in the increased available volume to be visible in surveys. This
``brightest observable cluster in a finite volume'' statistic is shown
both for a universe of trivial topology and for a universe of non-trivial
topology with a fundamental polyhedron diameter, for illustration purposes,
of 600$h^{-1}$~Mpc. 

If RXJ~1347.5-1145 did not exist, previous astrophysical object based
limits on the diameter of the fundamental polyhedron 
(e.g., \cite{LLL96}~1996) would be consistent with the existence
of the other bright clusters represented. One might be encouraged
to search for signs of identity among the four $z>0\.2$ clusters
shown in Fig.~\ref{f-Lxplot}, or to look for ghosts of these objects 
at higher redshifts. However, RXJ~1347.5-1145 does exist, and its
discovery at a moderately high redshift does not imply that clusters
were brighter in the past, it is simply consistent with being a rare
cluster requiring a large search volume, 
under the assumptions of a trivial topology of the Universe and 
no significant luminosity evolution!

Nevertheless, we do consider some possibilities of ghost images of
RXJ~1347.5-1145 below.

\subsection{Is Coma a Ghost of RXJ~1347.5-1145?}
Could the Coma cluster be a low redshift image of RXJ~1347.5-1145?
This would obviously be of tremendous usefulness in understanding both cluster
and galaxy evolution, given the huge time lag between the two images.

Both clusters have two large ``central dominant'' (cD) galaxies
at their centres; the total masses are identical within the uncertainties 
[\cite{Briel92}'s (1992) ROSAT mass estimate for Coma 
implies a total mass of $(4-10)\e{14}M_{\sun}$ to $1h_{50}^{-1}$~Mpc)
and $(10-20)\e{14}M_{\sun}$ to $3h_{50}^{-1}$~Mpc; \cite{Sch96} give
$5\.8\e{14}M_{\sun}$ and $17\e{14}M_{\sun}$ for 
RXJ~1347.5-1145 to the same radii respectively];
and the gas fraction to a large radius appears lower in Coma, but
consistent within the uncertainty.

However, the much preciser X-ray fluxes and core ($r<500h_{50}^{-1}$~kpc) 
surface brightnesses are much higher for RXJ~1347.5-1145 than for Coma.
If the ghost of RXJ~1347.5-1145 
were to lie as close to us as Coma, without any reduction in the
bremsstrahlung luminosity of the hot (and cooling) gas,
then it would have a $0\.1-2\.4$~keV X-ray
flux of $3\.4\e{-9}$~erg/s/cm$^2$, which is 
six times as bright as the Perseus galaxy cluster 
and ten times Coma itself. 
For RXJ~1347.5-1145 and Coma to be ghost images
of one another, 
this would require not only removal 
of the cooling flow, but also some means of further reducing the hot gas
emission by a factor of about ten.
Moreover, the core gas would have to be radically redistributed in the 
$\approx 2\.6h_{50}^{-1}$Gyr (for $\Omega_0=1, \lambda_0=0$) time lapse
between the observed two emission epochs. 
For this to be done, as suggested above,
by collision with another cluster, this second cluster should be visible close 
to RXJ~1347.5-1145 in the ROSAT pointed observations or should be
detectable in moderately deep optical imaging over a 10--15$'$ field
around the cluster.

So it seems difficult to see how Coma could physically be identical
to RXJ~1347.5-1145.

\subsection{Other Candidates for a Ghost of RXJ~1347.5-1145}
The fact that the redshifts of RXJ~1347.5-1145 and CL~09104+4109 are similar 
provides a time scale argument against these two clusters being physically
identical.

Since the redshifts are close, 
we in fact see CL~09104+4109 at an epoch only 35$h^{-1}$~Myr 
(for $\Omega_0=1, \lambda_0=0$) later than RXJ~1347.5-1145, since the
light travel times are indicated by the redshifts, irrespective of topology.
A very IR-luminous quasar is seen in CL~09104+4109, but a strong IR source 
has not been found in RXJ~1347.5-1145. If the ordering here were the opposite,
an estimate of the lifetime of the IR emission $\gg 35$~Myr would make 
it hard to understand why the emission is not seen in RXJ~1347.5-1145.
However, RXJ~1347.5-1145 is the earlier image, so we can only consider
the probability that two nearly simultaneous snapshots of the cluster
happen to occur before and during the period of IR emission.

What is the likely lifetime of the far-IR emission? 
Starlight from stars created by the process causing the quasar, 
e.g., galaxy merging, reradiated in the far-IR is a likely candidate.
A typical dynamical time scale for such an event, supposing this involves
two typical large galaxies
of $10^{12}M_{\sun},$ is about $200$~Myr
(e.g., \cite{ScoSoi91}~1991), while the massive stars formed during
this period would remain on the main sequence for about this much
time afterwards. \cite{Cavpad88}~(1988) estimate that quasar lifetimes
are likely to be at most about a Gigayear. In either case, in the case
of physical identity of RXJ~1347.5-1145 and CL~09104+4109, it would seem
to be about a 10\% coincidence that we happen to see one ghost 
image just before the event started and one during the event. 

Most of the other candidate ghost images of RXJ~1347.5-1145 listed
in Table~\ref{t-clusters} seem to be of lower luminosity and are at
lower redshifts, requiring a reduction in $L_X$ with time in order
to be identical with RXJ~1347.5-1145, which as discussed above is
physically unlikely. On the other hand, A2163 is at lower redshift
than RXJ~1347.5-1145, but is considerably more massive, making it
too an unlikely ghost candidate.

\subsection{A Ghost in the Plane?}

Could a ghost of RXJ~1347.5-1145 be hiding in the galactic plane? 
The fraction of the soft X-ray flux absorbed by the Galaxy
varies from 10--20\% at the galactic poles to 95\% on the plane
(assuming a galactic plane column density of $n_{H}=3\e{22}$~cm$^{-2}$
for $|b^{II}|=0\deg$). 
This problem is even
worse if one considers the detected ROSAT count rate which is
predominantly at 1~keV where the detected count rate is
down by 98\% on the plane.
At a galactic latitude of
$|b^{II}|=20\deg$ the fraction of absorbed flux 
is about 50--60\% (with a large variance), so
studies in soft X-rays (e.g. \cite{Ebel93}~1993) are
limited to the two thirds of the sky at higher latitudes.

The existence of a ghost in the plane is therefore a possibility,
which cannot easily be excluded.
However, this is a fundamental limitation to any search for individual 
ghost images, which affects all other candidate objects.

One future possibility would be for the 
ABRIXSAS satellite to perform an all-sky survey in hard ($E>3$~keV) 
X-ray bands, in which the galactic plane is substantially more transparent. 
However, in this case, since it would be difficult to confirm any
hard X-ray detected candidate clusters at other wavelengths, a physical
test for distinguishing clusters from other sources in
the infrared or using the hard
X-ray data alone (e.g., from surface brightness profiles) 
would need to be established from the sources at high
galactic latitudes. Whether or not this is feasible remains to be seen.


\subsection{A Candidate Multi-connected Manfold}

While the ``local physics'' arguments just presented 
argue against either Coma or CL~09104+4109 
being ghost images of RXJ~1347.5-1145, 
it is quite exciting to note nevertheless 
that examination of the three-dimensional
positions of these bright clusters
yields a geometrical pattern indicative of a candidate multi-connected
manifold; specifically, a hypertorus, of which two side lengths 
are just under $1000$~kpc.
RXJ~1347.5-1145, the Coma cluster, and the cluster CL~09104+4109 
(\cite{Hall97}~1997)
form nearly a right angle ($\approx 88\deg$) with arms of nearly identical 
length ($970h^{-1}$ and $960h^{-1}$~Mpc respectively) for 
$\Omega_0=1, \lambda_0=0$ curvature. 

Of course, 
the CMB analyses cited above specifically concentrate on the case of 
a hypertoroidal universe, 
since this is the simplest of non-trivial topologies 
possible, so it is unlikely that a hypertorus on the scale required for
identity of the three clusters has managed
to escape attention so far. However, this geometrical configuration is
one which has already been searched for among other objects
(e.g., quasars, \cite{Fag87}~1987), and was found among a very small
number of candidate objects, so it is certainly interesting to consider
independently of the CMB analyses.

\FigQobs 
\FigQsim 

\tabbas

The mapping from the universal covering space to a single copy of the
fundamental polyhedron, which in this case is assumed to be a cube 
(although other possibilities for the third dimension could be considered),
is very simple. By forcing the three clusters to form an exact right angle
of arms (axes) of identical length, and choosing the third axis to be at
right angles to the first two and of the same length, the
transformation from the universal covering space to the fundamental
polyhedron is simply 
\begin{equation}
{\bf r}' = [{\bf r}.{\bf e}_1 (\mod\, L),\; {\bf r}.{\bf e}_2 (\mod\, L),\;
{\bf r}.{\bf e}_3 (\mod\, L)]
\label{e-hyper}
\end{equation}
where ${\bf r}$ is the (three-dimensional) 
position of any astrophysical object, 
$({\bf e}_1, {\bf e}_2, {\bf e}_3)$ is the orthonormal basis listed 
in Table~\ref{t-basis}, $L$ is the side length of the fundamental cube
and ${\bf r}'$ is the object's position translated into the fundamental
cube and expressed in the coordinate system of the 
orthonormal basis.\footnote{Note to Fortran users: the Fortran $\mod(a,b)$
function should be modified for use with negative values of $a$.}

Readers can easily use this transformation to check for themselves whether
or not catalogues of objects at large redshifts transformed into the 
(candidate) fundamental cube happen to coincide with one another---as should
be the case if the transformation is due to a genuine physical identification.

An example of such an application is shown here, using a list of 
5007 quasar positions
(from the NASA/IPAC Extragalactic Database, NED), 
the highly luminous clusters discussed above, 
a few large superclusters, and the cold spots in the CMB tentatively
attributed to density peaks (\cite{CaySm95}~1995). The positions of
all these objects are transformed to the fundamental cube and plotted
in Fig.~\ref{f-qobs}. Rotation of the fundamental cube by an arbitrary
angle (i.e., no physical justification) gives a control sample
in Fig.~\ref{f-qsim}.

If identical quasars were seen as several different ghost images, 
then more close pairs should be seen in the map based on the cluster-derived
fundamental cube than that based on the arbitrarily rotated fundamental
cube. An excess of close pairs is not obvious to the eye, 
and a two-point correlation function (using 10 different control samples) 
confirms the lack of any statistical
difference.

It could be the case that the angles of the fundamental cube are not
perfect right angles, or that the sides are of slightly unequal lengths,
but in that case the quasars should still follow 
large scale structure, on a scale of
about $50-150h^{-1}$~Mpc (e.g., \cite{deLapp86}~1986; \cite{GH89}~1989;
\cite{daCosta93}~1993; \cite{Deng96}~1996; \cite{Einasto97}~1997). 
Again, the two-point correlation function shows no quantitative difference
between structure in the case of the candidate fundamental cube and 
the control cases.

Note that the large scale variations in quasar density are due to
the quasar catalogue containing a variety of observational catalogues
over small areas of the sky, as well as large solid angle surveys. This
is not a sign of multi-connectedness.

The group transformation indicated in Eq.~\ref{e-hyper} and 
Table~\ref{t-basis} could also be applied to search for other 
copies of the ``corners'' of the fundamental polyhedron, each 
position at which another image of the (hypothetically single) cluster 
RXJ~1347.5-1145/Coma/CL~09104+4109 should be visible. Since Coma
is nearly at the location of the observer, many of these images 
would be hidden behind one another. However, the four images at
\begin{eqnarray}
{\bf r}_A &\equiv& \rComa - {\bf r}.{\bf e}_1 
= (0\.392, 1^h59^m, +18\deg),  \nonumber\\ 
{\bf r}_B &\equiv& \rComa - {\bf r}.{\bf e}_2
= (0\.395, 20^h32^m, -41\deg),  \nonumber \\ 
{\bf r}_C &\equiv& \rComa + {\bf r}.{\bf e}_3 
= (0\.406, 7^h12^m, -43\deg) \mbox{\rm \ and} \nonumber\\
{\bf r}_D &\equiv& \rComa - {\bf r}.{\bf e}_3 
= (0\.434, 18^h31^m, 46\deg) 
\end{eqnarray}
[written as ($z$, $\alpha,$ $\delta$)] would not be hidden. 
Firm constraints on the existence or non-existence
of clusters at these positions, within uncertainties 
similar to the difference of the observed 
cluster triplet from an exact right angle
of equal arm lengths (i.e., of order 1\%), should be relatively easy to 
obtain observationally. Again, our closeness to Coma implies that 
the images should be seen at similar redshifts to those of 
RXJ~1347.5-1145 and CL~09104+4109, so the possibility that they 
are much fainter than these two clusters would require corresponding 
changes in luminosity over very short time intervals.

The lack of obvious candidate ghost images at these positions 
in the ROSAT All-Sky Survey (RASS) provides
an additional argument against the candidate topology suggested. 

However, for other candidate topologies which retain 
the identification of the three observed clusters, but are not
simply $T^3$ with three equal fundamental lengths, 
ghost images would only have to be seen at 
positions ${\bf r}_A$ and ${\bf r}_B,$ the near-``antipodes'' 
of RXJ~1347.5-1145 and CL~09104+4109.
RASS exposure in the direction of 
${\bf r}_A$ is not deep enough to exclude a counterimage cluster and
RXJ~203150.4-403656 is a plausible candidate at ${\bf r}_B.$ 
So, spectroscopy to determine the redshift of the latter 
and an optical (or X-ray) 
search for a cluster within $2\deg$ of ($1^h59^m, +18\deg$) 
and within $\delta(z) \ltapprox 0\.005$ of $z=0\.392$ would be needed to
rule out variants on the candidate manifold suggested.

A final argument against the candidate manifold is the rarity of 
the cluster RXJ~1347.5-1145/\-Coma/\-CL~09104+4109. According to the 
hypothesised manifold,
this would be (historically) the brightest cluster 
in the Universe---and would happen to be just next door (70$h^{-1}$~Mpc away). 
The probability of the historically brightest cluster 
being this close to us in the
Universe of total volume $(962 h^{-1}$~Mpc$)^3$
is simply $P= [4\pi(70)^3/3]/(962)^3 = 0\.0016.$ 

For a simply-connected
topology, the brightest cluster is RXJ~1347.5-1145 in its observed
manifestation. As shown in Fig.~\ref{f-Lxplot}, the probability of 
the occurrence of RXJ~1347.5-114 at this distance is quite high.
Even the anthropic argument would not seem to be much help here.

\section{Conclusion} \label{s-concl}
Highly luminous X-ray clusters are a robust probe for finding ghost
images of astrophysical objects which would reveal the 
possible non-triviality 
of the topology of the Universe. This is because 
observational statistics indicate that these 
clusters are likely to increase (or at a minimum retain the same) 
X-ray 
luminosity as time increases; while theoretically, gravity and conservation
of matter imply that it is hard to see how the situation could be 
different for individual clusters.

Observations of successively brighter (more massive) rich clusters at
higher redshifts implies successively greater volumes in which the
Universe must be simply connected, while observation of the most X-ray 
bright (most massive) cluster at a redshift well below that to 
which virialised clusters are (eventually) discovered would be a clue
to a multi-connected Universe. In the latter case, this brightest cluster
would have ghost images among the population at higher redshifts.

The apparent lack of ghosts of RXJ~1347.5-1145 implies a lower 
limit to the size
of the fundamental polyhedron, $\beta,$
of about $1000h_{50}^{-1}$~Mpc (for $\Omega_0=1, \lambda_0=0$). This limit
could be doubled in size, without having to observe to fainter flux
limits than those of the ROSAT All-Sky Survey, by a survey
through the galactic plane in hard X-rays, if some means of 
confirming which sources are clusters could be found (e.g., by surface 
brightness profiles combined with spectral shapes).

The consideration of highly luminous X-ray clusters 
has indicated what is geometrically 
an exciting candidate identification of ghost images.
RXJ~1347.5-1145, the Coma cluster and the cluster CL~09104+4109 
together form what is nearly a right angle of nearly equal arm lengths,
just what would be expected for a hypertoroidal geometry (for flat 
curvature). The inferred transformation from the covering space to
the fundamental cube is presented in Table~\ref{t-basis} and 
Eqn~\ref{e-hyper}.
However, the local physical properties of these three clusters
do not seem to support identity; 
CMB constraints against hypertoroidal topologies have been well studied;
a mapping of $\approx 5000$ 
quasar three-dimensional positions into 
the implied fundamental cube of the hypertoroidal manifold does not
support identity; not all of the expected four extra ghost images of 
RXJ~1347.5-1145 and CL~09104+4109 are obvious; and the probability 
that we are as close as we are to the historically brightest cluster 
in the Universe RXJ~1347.5-1145/Coma/CL~09104+4109 would be
only $0\.16\%.$

Nevertheless, observations to determine the redshift of 
RXJ~203150.4-403656 and to search for a cluster
within $2\deg$ of ($1^h59^m, +18\deg$) 
and $\delta(z) \ltapprox 0\.005$ of $z=0\.392$ would be useful to
rule out (or detect!) variants on the candidate manifold suggested.

\section{acknowledgements}
We thank the (anonymous) referee for a careful reading of the paper
and useful suggestions.
This work has partly been carried
out under a Centre of Excellence Foreign Visiting Fellowship of the 
Japanese Ministry of Education (Monbush${\rm \overline{o}}$) and has
made use of the NASA/IPAC extragalactic database (NED)
which is operated by the Jet Propulsion Laboratory, Caltech, under contract
with the National Aeronautics and Space Administration. ACE would
like to thank the Royal Society for support.



\end{document}